# Increasing Gender Balance Across Academic Staffing in Computer Science  - case study

Dr. Susan McKeever, Dr Deirdre Lillis

*School of Computer Sciemce, Technological University Dublin*

**Abstract**

As at 2019, Technological University Dublin[*] Computer Science is the top university in Ireland in terms of gender balance of female academic staff in computer science schools. In an academic team of approximately 55 full-time equivalents, 36% of our academic staff are female, 50% of our senior academic leadership team (2 of 4) are female and 75% of our School Executive are female (3 of 4), including a female Head of School. This is as a result of our seven year SUCCESS programme which had a four strand approach: Source, Career, Environment and Support. The *Source* strand explicitly encouraged females to apply for each recruitment drive; *Career* focused on female career and skills development initiatives; *Environment* created a female-friendly culture and reputation, both within the School, across our organisation and across the third level sector in Ireland and *Support* addressed practical supports for the specific difficulties experienced by female staff. As a result we have had 0% turnover in female staff in the past five years (in contrast to 10% male staff turnover). We will continue to work across these four strands to preserve our pipeline of female staff and ensure their success over the coming years in an academic and ICT sector that remains challenging for females.

## 1. Introduction

The SUCCESS initiative in TU Dublin Computer Science, started in 2012, has increased the numbers, improved the experience and impact of female academic staff - such that they then in turn provide role models and supportive learning environments for our female students. The purpose of the SUCCESS initiative is two-fold (i) To achieve gender equality by increasing the number of female academic staff in our School and (2) to support female academic staff to achieve their career goals in a supportive work environment. These are not trivial aims, in an  context where average female staff levels in Computer Science in Ireland is 19% and where the average in the broader ICT sector is 10%.

---

[*] Technological University Dublin (TU Dublin) was established on 1st January 2019, arising from a merger of Dublin Institute of Technology, Institute of Technology Tallaght and the Institute of Technology Blanchardstown. DIT was a member of Informatics Europe.

## 2. SUCCESS Initiative Description

Ireland, as with most countries in Europe, suffers from a lack of females in both the IT sector, and more worryingly, in the pipeline of female school leavers into Computer Science programmes at third level. This is compounded by difficulties in achieving gender equality in academic staff. In Ireland, we have a national average of 19% females on academic staff in Schools of Computer Science. When SUCCESS started in 2012, the analysis identified a number of problems for recruiting and retaining female staff, problems which still exist across Ireland today [1] [3]. These include
- Advertised academic posts receive far fewer numbers of female applicants than male
- Females do not progress in equal proportions as men to management or professor grades and take on heavy loads of administration management roles, leaving less time for career enhancing research;
- Female staff find it difficult to return after maternity leave with new teaching loads and gaps in research output;
- The networks and culture of academic institutions are created and populated by males.

**SuCcEsS** focuses on addressing these problems (1) Source: Increase the number of female academic staff applying for academic positions (2) Career: Run female-focussed skills development initiatives (3) Environment: Create an environment of support and encouragement for female staff (4) Support: Implement practical supports for female staff.

### 2.1. Source

This strand has a simple focus - gender balance in our recruitment. When a vacancy arises, we identify female candidates external to our university/School and encourage them to join our team, using our networks to attract and target suitable female staff. In the past year alone for example, we have recruited a departmental head, specifically encouraged to apply for the role (but with no further advantage) and an engineering lecturer converted to Computer Science.

### 2.2. Career

Female staff have in the past been more reluctant than male staff to take on career enhancing roles, such as new research opportunities or new management roles. We run a number of initiatives that are aimed at boosting our female staff skills, confidence levels and career ambition.
- All staff have an annual performance review, where their career aims and performance are reviewed and aligned to the strategic goals of TU Dublin. This Performance Management Development System (PMDS) allows female staff to articulate any particular difficulties or aims which the School can support them on. PDPs are conducted by School Executives (the majority of whom are female also).
- The School funds female staff to participate in the Aurora leadership programme for female academics[†], run by the Leadership Foundation in the UK and also TU Dublin leadership development initiatives. 33% of our female staff have taken part, with several now acting as mentors for the programme.

---

[†] https://www.lfhe.ac.uk/en/programmes-events/programmes/women-only/aurora/

- The School funds early career researchers through a number of supports which include support for conferences, inclusion on PhD supervision schemes and seed-funding of research proposals.
- The School funds a part time PhD scheme for any staff who wish to undertake a PhD by covering fees and workload allowances. This is particularly relevant and important for female staff who are more likely to be recruited from an industry background.
- Female staff are encouraged to mentor more junior female staff, in particular on co-supervision in PhD students, and co-funding opportunities.

*2.3. Environment*

Gender equality permeates everything we do and stand for as a School, we lead by example and have a reputation for excellence within TU Dublin and across the third level sector in Ireland.

- *Within our School*: We offer role models to our female staff via our female school managers. For example, our Head of School, Dr. Deirdre Lillis is a board member of the Higher Education Authority and a member of the Senate of the National University of Ireland. She participates actively in the development of national policy developments for gender equality in higher education, including the new Senior Academic Leadership Initiative[‡] and the Gender Equality Action Plan for Higher education in Ireland[§]. For students, we run the ESTEEM student mentoring scheme whereby every female student in our School is teamed with a female staff member and mentored by an industry role model. This initiative which has been identified as best practice in gender balance activities by the UK-based Athena Swan Equality Unit[**]. These activities enhances our reputation and culture.
- *Within TU Dublin:* Our School has played a key part in establishing and participating in The Women Leaders in Higher Education (WLHE[††]) Network. The network, established in 2016, empowers and enables women in all roles (management, academic, technical and administrative) in TU Dublin, to support their personal and professional development and career advancement. The WLHE Network holds formal events, with invited speakers from a wide range of backgrounds including media personalities, business leaders, academics, and politicians.
- *Within the national third level sector*: Our School led the development of a national network for academics in third level Computer Science called **Ingenics**[‡‡] (The Irish Network for Gender Equality in Computing). This was set up 2017 is attended by 17 out of 19 Irish third level Computer Science schools - and is proving to be an invaluable and powerful network of insight, information sharing, joint projects, networking and friendships.
- TU Dublin obtained its first Athena Swan award in 2018. The international award recognises commitment to advancing the careers of women in science, technology, engineering, maths and medicine (STEMM) employment in higher education and research. Our focus now is to expand Athena for its next stage of School level approval.

---

[‡] https://www.education.ie/en/Press-Events/Press-Releases/2018-press-releases/PR181112.html
[§] https://hea.ie/assets/uploads/2018/11/Gender-Equality-Taskforce-Action-Plan-2018-2020.pdf
[**] https://www.ecu.ac.uk/equality-charters/athena-swan/
[††] http://www.dit.ie/wlhe/
[‡‡] https://www.linkedin.com/groups/13676439/

*2.4. Support*

The final strand of SUCCESS is practical supports for female staff.  Although simple, these are critical in addressing the practical problems that female staff in an academic environment. Given the age profile of our female staff, we placed particular emphasis on supports after maternity level. These include

- Teaching modules prior to maternity leave are re-allocated to these staff on their return so that they are not disadvantaged by heavy preparation workloads.
- Females can be assigned flexible project work if they return mid-semester.
- Timetabling is done with consideration of family commitments such as consideration of early morning starts or evening workloads.  Flexibility on home working is provided as suitable.

## 3. Evidence of Impact of the SUCCESS Initiative

In our experience, there are no easy or quick solutions to addressing the issue of recruitment numbers and career development  for female academics in the Computer Science sectors. Our SUCCESS initiative has been a sustained, consistent programme of supports and activities.  We measure our impact in this area by answering the following questions:

- Are we recruiting female staff in sufficient numbers?
- Are female academics succeeding in our School?
- Are we retaining female academics in our School?

*Source: Are we recruiting female staff in sufficient numbers?*
In a survey of Irish third levels with an 80% response rate from 17 universities and Institutes of Technology in our Ingenics network (Figure 1), we have the third highest proportion of academic staff that are female. In comparison to universities only, we have the highest proportion of female staff - at 36%. Our experience in 2012 was that less than 10% of applicants to entry-level academic posts were female. Through consistent efforts this has increased to 25% on average, to 50% in some competitions and has led to hiring 5 new females in the past 5 years (out of 12 vacancies).

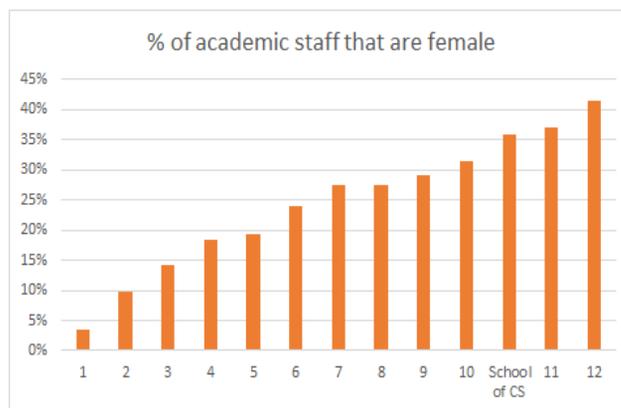

**Figure 1 - proportion of female staff in Schools of  Computer Science in the third level sector in Ireland, across universities and institutes of technology.**

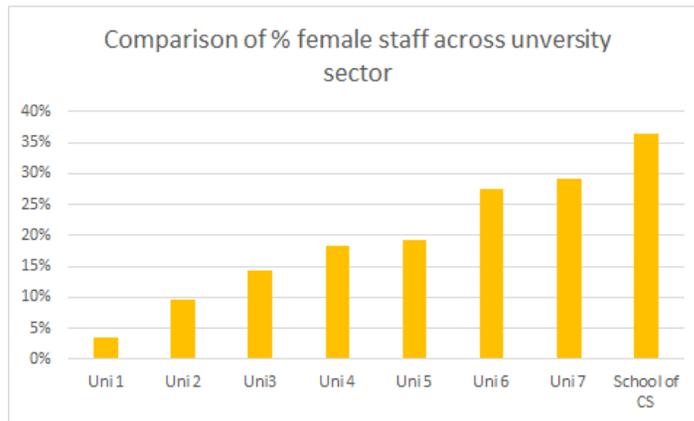

**Figure 2 - proportion of female staff in Schools of Computer Science in university sector, showing School of Computer Science TU Dublin in highest position**

*Career: Are female academic succeeding in our School?*
In addition to increasing our female numbers, it is critical that we have proportional representation at management decision making level - ensuring supportive decision for all staff. As shown in Figure 3, the percentage of females on our (i) School Executive team has increased from 0% in 2008 to 75% in 2019 (4 positions) and (ii) on our senior academic leadership team from 0% in 2008 to 50% in 2019 (4 positions) and (iii) our first full Professorship in this School was female. We have supported 2 female staff to complete their PhDs through timetable workload allowances. Our female staff also play leading roles at international, national and institutional level on a wide range of initiatives including national research centres (Science Foundation Ireland and the national industry-led Data Analytics Centre CeADAR), EU international curriculum development projects ([www.hublinked.eu](www.hublinked.eu), [www.getm3.eu](www.getm3.eu)), Boards of national policy agencies, the Higher Education Authority, the Senate of the Irish Universities, the Royal Irish Academy and the Ingenics network. They also play leading roles within TU Dublin on the Academic Council, College Boards, quality enhancement initiatives, the ESTEEM mentoring programme, community projects, programme development initiatives and the Women's Leaders in Higher Education Initiative. Our policies of developing female staff have had impact.

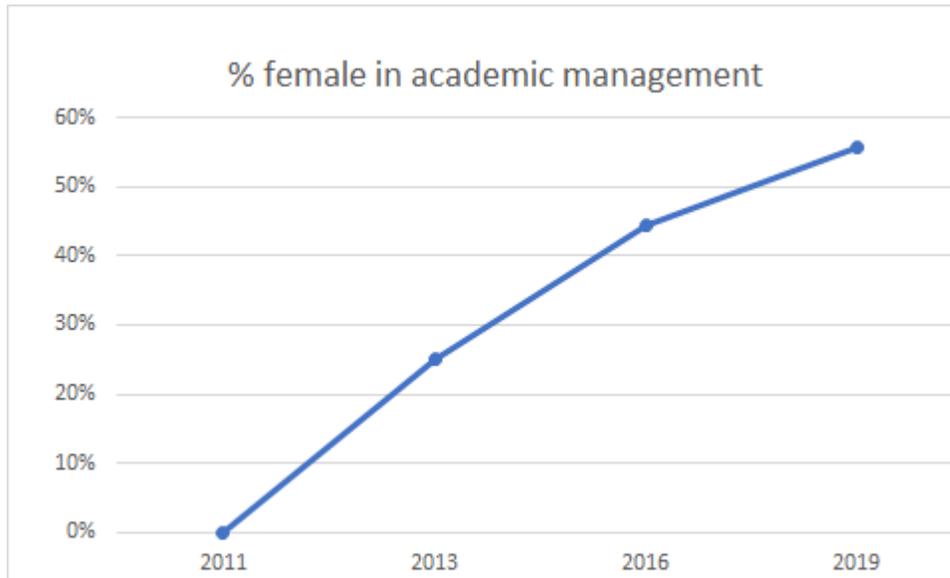
**Figure 3 - Increase in female management staff since the start of SUCCESS initiative**

*Environment: Are we retaining female staff in our School?*
For staff retention, we note that in the past five years, we have had no female staff turnover, compared against 10% turnover amongst male staff.  While it may be simplistic to attribute all of this to SUCCESS, this indicates that females are satisfied to remain in our School. The influence of female staff can be seen on our strategic planning and development processes as research has shown that females are often more motivated by impact and the social aspects of technology. The theme for our strategic plan developed in 2019 focussed on how the School could help achieve the UN Sustainability Goals for example. Led by one of our female academics, we submitted an Erasmus+ Strategic Partnership called Ethics4EU, with Informatics Europe as a partner, which aims to learning resources and assessment strategies for the ethical use of technology which can be embedded in all modules of our programmes.

In our School Review in 2017, our external peer review panel noted the high satisfaction rates amongst staff with the School, across both male and female staff. They also recommended that we be the first School to be put forward for Athena Swan Accreditation from TU Dublin as a result of our work on gender equality.

## 4. Conclusion

We are national leaders in Ireland on our gender equality initiatives and are proud of our record on supporting female academics in our School under SUCCESS. While we have achieved many of the goals we set out to achieve, this will be a work in progress until we have true gender equality for both staff and students.

## 5. Gallery

than-men-in-advancing-to-the-highest-academic-positions-in-ireland/

[2] Higher Education Authority Ireland, 2018, Higher Education Institutional Staff Profiles by Gender, published July 2018, available at https://hea.ie/assets/uploads/2018/01/Higher-Education-Institutional-Staff-Profiles-by-Gender-2018.pdf

[3] Dublin Institute of Technology Athena Swan Bronze Application, 2018.

## Appendix A. Gallery

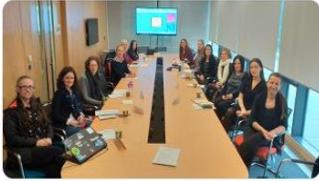

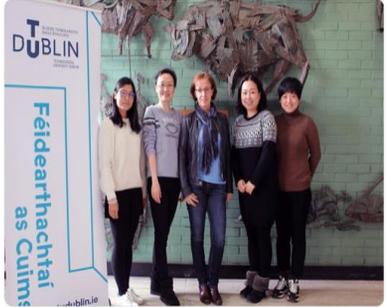

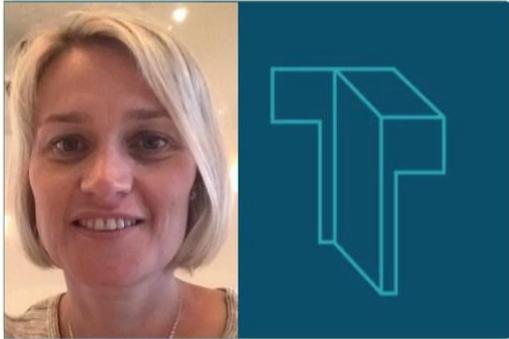

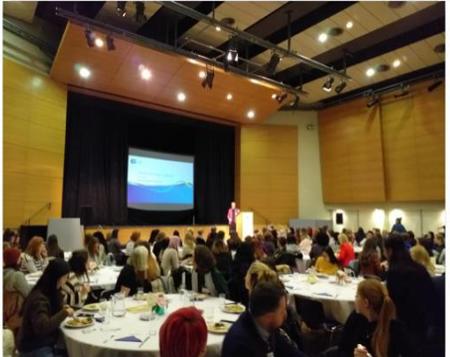

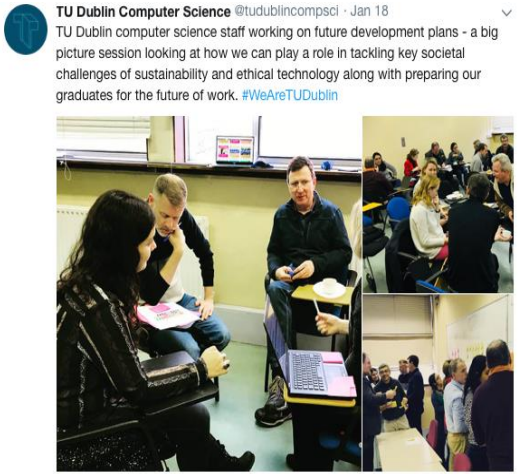
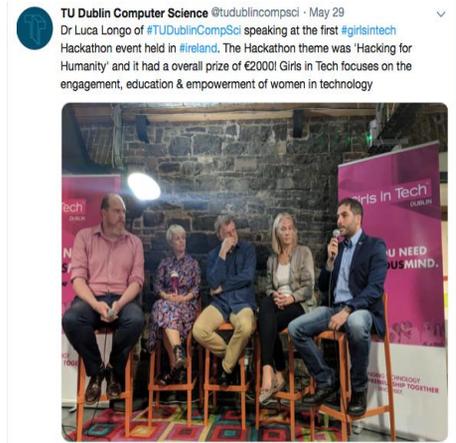
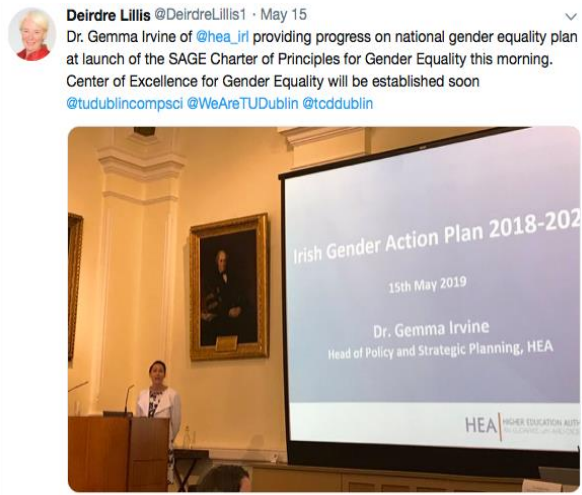
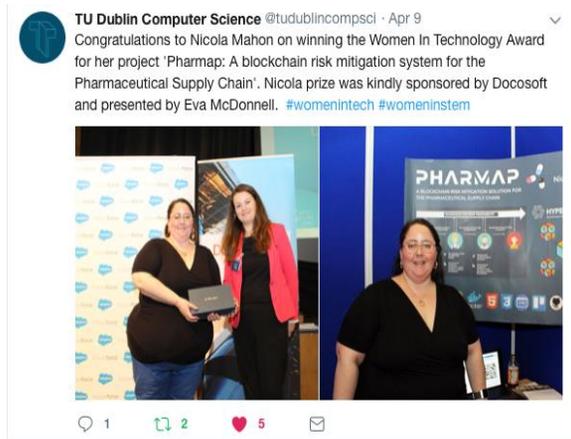
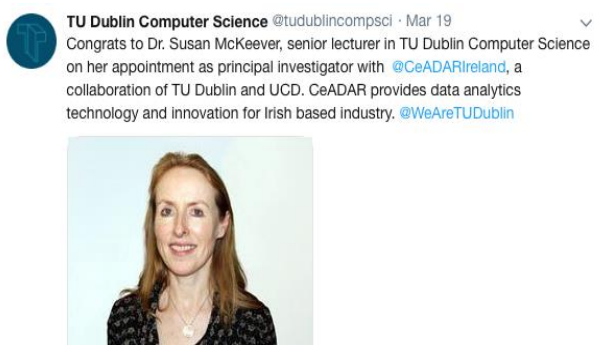
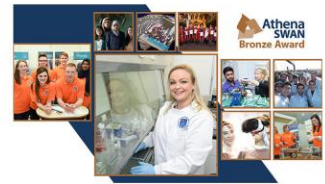

### Dr Deirdre Lillis appointed to Senate of NUI

Posted: 22 January, 2019

The Minister for Education and Skills, Joe McHugh TD, has appointed Dr Deirdre Lillis, TU Dublin, to the Senate of the National University of Ireland (NUI) for the 2019-2024 term.

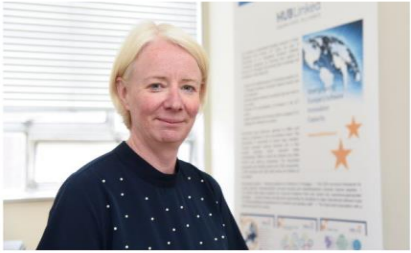

The NUI, and its member universities and colleges, promotes the National University of Ireland nationally and internationally through scholarship and the advancement of higher education and the cultural and intellectual life of Ireland. Dr Lillis is one of four Government appointees to the Senate, including Professor Patrick Clancy, Dr Barbara Doyle Prestwich and Mr John Hurley.

Since 2008, Dr Lillis has been Head of the School of Computer Science in TU Dublin's City Campus. The School is the largest provider of computer science graduates to the ICT sector in Ireland. She has led an integrated internationalisation strategy, working with partners in Europe, Korea and China and has secured significant European funding for projects in university-industry innovations and young talent management.

### TU Dublin Partners in Three New SFI Centres for Research Training

Posted: 8 March, 2019

Minister for Business, Enterprise and Innovation, Heather Humphreys TD and Minister of State for Training, Skills, Innovation, Research and Development, John Halligan TD have announced an investment of over €100 million in six new SFI Centres for Research Training Centres.

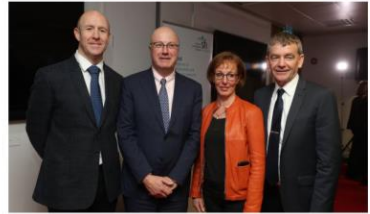

*Prof. John Kelleher (TU Dublin), Prof. Mark Ferguson, Director General at Science Foundation Ireland and Chief Scientific Adviser to the Government of Ireland, Professor Sarah Jane Delany (TU Dublin) and Prof. Max Ammann (TU Dublin)*

TU Dublin academics have been named partners at three of the six new SFI centres, which will provide training for 700 postgraduate students in the areas of nationally and internationally identified future skills needs of digital, data and ICT.

**Professor Sarah Jane Delany - SFI Centre for Research Training in Machine Learning**

From self-driving cars to chess-playing computers, from autonomous booking agents to automated financial trading systems, applications of Artificial Intelligence (AI) are having massive impacts in ever-growing aspects of our lives.

---

[i] https://www.informatics-europe.org/awards/40-awards/minerva-informatics-equality-award/winners/503-minerva-winner-2019.html